\pdfoutput=1
\documentclass{article}
\usepackage{frascatiphys,graphicx,subfigure}
\usepackage{amsmath}
\usepackage{amsthm}
\usepackage{booktabs}
\usepackage{epsfig}
\begin{document}
\title{\textbf{Design, R\&D and status of the crystal calorimeter for the Mu2e experiment}}
\author{
Raffaella Donghia on behalf of the Mu2e calorimeter group\\
{\em Universit\'a degli Studi Roma Tre, Rome, Italy}, 
\\
{\em Laboratori Nazionali di Frascati - INFN, Frascati, Italy } \\
}
\maketitle
\baselineskip=11.6pt
\begin{abstract}
The Mu2e Experiment at Fermilab will search for coherent, neutrinoless conversion of 
muons into electrons in the field of a nucleus, with a sensitivity improvement of a factor of 10$^4$ over previous experiments. Such a charged lepton flavor-violating reaction probes new physics at a scale inaccessible with direct searches at either present or planned high energy colliders.  

The conversion electron is  mono-energetic with an energy slightly below the muon rest mass. 
If no events are observed in three years of running, Mu2e will set a limit on
the ratio between the conversion rate and the capture rate, R$_{\mu e}$, of 
$\leq 6 \times 10^{-17}$ (@ 90$\%$ CL).

In this paper, the physics motivation for Mu2e and the current status of the  electromagnetic calorimeter project are briefly presented. 
\end{abstract}
\baselineskip=12.pt
\section{Introduction}
The Mu2e experiment at Fermilab\cite{tdr} will search for the charged lepton flavor violating (CLFV) 
process of muon conversion in an $^{27}$Al nucleus field, $\mu + N(Z, A) \rightarrow e + N(Z, A)$. 
No CLFV interactions have been observed experimentally yet. 
The current best limit on $\mu-e$ conversion has been set by SINDRUM II experiment.\cite{sindrum} 
Mu2e intends to probe 4 orders of magnitude beyond the SINDRUM II sensitivity, measuring  the 
ratio, $R_{\mu e}$, between the conversion rate to number of muon captures by Al nucleus:
\begin{equation*}
R_{\mu e} = \frac{\mu^- \thinspace N(Z,A) \rightarrow e^- \thinspace N(Z,A)}{\mu^- \thinspace N(Z,A) \rightarrow\nu_{\mu} \thinspace N(Z-1,A)} < 6 \times 10^{-17},~(@~90\% CL)
\end{equation*}

The signature of this neutrinoless conversion process is a monoenergetic electron, 
with an energy slightly lower than the muon rest mass, $\sim$~104.96 MeV. 
In order to achieve our goal, a very intense muon beam ($\sim$ $10^{10}$~Hz) has to 
stop on an aluminum target and a precise momentum analysis has to be performed. 

In the Standard Model (SM) the expected rate is negligible 
(BR $ \sim~10^{-54}$), so that, observation of 
these processes should be crucial evidence of New Physics beyond the SM.\cite{clfv_sm} 

\section{Calorimeter requirements}
The Mu2e calorimeter is designed to identify $\sim$~100~MeV electrons 
and to reduce the background to a negligible level. It is located inside a large superconducting 
solenoid, just behind the tracker, which complements it. Indeed, the calorimeter 
provides information about energy, timing and position to validate charged particles 
reconstructed by the tracker and reject fakes.
Moreover, the calorimeter has to perform a particle identification to distinguish muons from 
electrons. These tasks lead to the following requirements\cite{tdr}:
 an energy resolution around  5\% (5 MeV, at 100 MeV);
 a timing resolution better than 0.5 ns;
 a position resolution better than 1 cm;
 little deterioration for radiation exposures up to $\sim$~100 krad in the hottest 
 region and for a neutron flux equivalent to $10^{12}$~MeV/cm$^2$;
Moreover, the Mu2e calorimeter must operate in 10$^{-4}$~Torr internal pressure within the 1 T magnetic field. 
This implies the use of solid-state photodetectors and of electronics (HV and FEE) 
immune to the presence of the magnetic field.

\section{Calorimeter design}
\begin{figure}[h]
\includegraphics[scale=0.50]{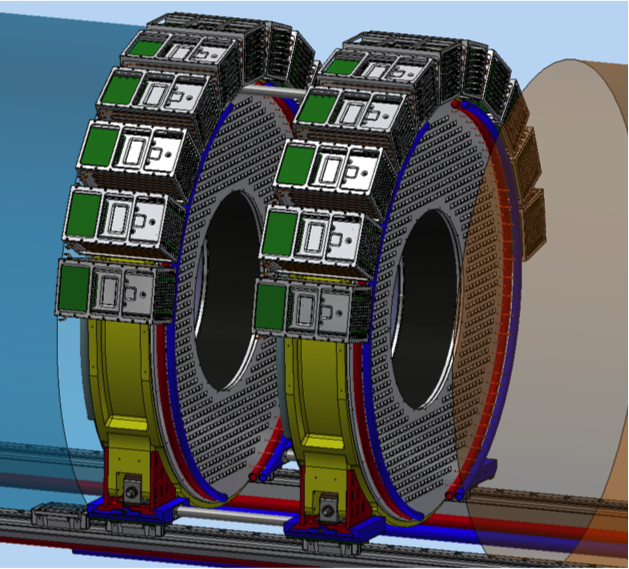}
\hspace{1.5 cm}
\includegraphics[scale=0.4]{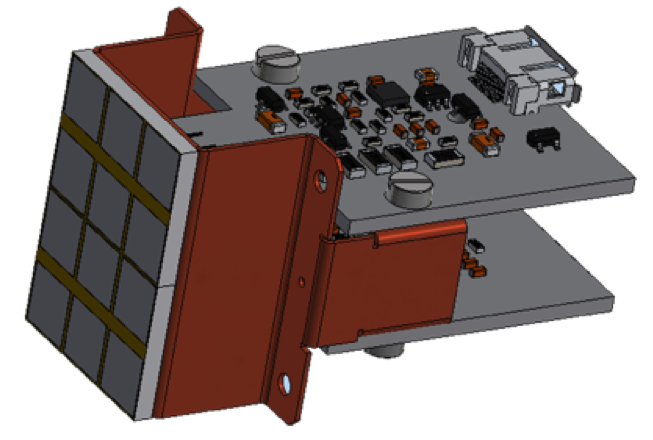}
\caption{\label{fig1} Annular disks structure of the Mu2e electromagnetic calorimeter (left). Layout of the 2 SiPMs coupled to each crystal, with the analog read out  electronics connected.}
\end{figure}

In the 100 MeV energy regime, a total absorption calorimeter employing a 
homogeneous continuous medium is required to meet the Mu2e requirements.
We decided to adopt a solution with two annular disks made by scintillating crystals, each readout using two 
solid state photon-counters.
Each disk (Fig.~\ref{fig1}, left) has an internal (external) radius of 374~mm (660~mm) 
and is filled with 674 ($34\times34\times200$)~mm$^3$ crystals. The two disks are 
separated by about half electron wavelength (70~cm). 
%

Due to the physical and geometrical constraints stated, crystals with high light output (LY), 
good light response uniformity (LRU $\geq$ 10\%),
fast signal ($\tau \leq$ 40~ns), 
radiation hard (with maximum LY loss below 40\%) and 
small radiation induced readout noise (below 0.6\%) are needed.

Different types of crystals have been considered: 
lutetium-yttrium oxyorthosilicate (LYSO), 
Barium Fluoride (BaF$_2$) and pure Cesium Iodide (CsI). 
In the CDR\cite{cdr}, the baseline calorimeter choice was LYSO crystals readout with APD and many tests were carried out for this option.\cite{roberto} A large increase price in 2013 made this option unaffordable, so that for the TDR\cite{tdr} we have opted for cheaper crystals such as BaF$_2$ and CsI. After a long R\&D program, we have finally selected undoped CsI crystals as baseline choice.\cite{raffa_elba}\cite{raffa_calor}

The CsI crystals readout is done by UV-extended silicon photomultipliers (SiPMs).
The requirement of having a small air gap between crystal and photodetector and the request of redundancy in the readout implies the use of custom devices. For the Mu2e experiment we have increased the transversal dimension of the CsI from ($30 \times 30$) to ($34 \times 34$)~mm$^2$ in order to accomodate two ($2 \times 3$) arrays of $6 \times 6$ mm$^2$ UV-extended SiPM. The samples already procured show a good PDE ($\sim$~30\% at 315~nm) with a gain greater than 10$^6$ at the operation voltage. 

Each SiPM is directly connected to the readout electronics (Fig.~\ref{fig1}, right) and to a dedicated board housing a transimpedence preamplifier with a settable gain $\times$15 or $\times$30, 2~V dynamic range and 15~ns rise time. This digital boards are housed into 11 crates (in the top of each disk) per disk with 20 differential channels per board. These boards are composed by a mezzanine board for input of SIPM signals and HV setting and a Waveform Digitizer section based on SmartFusione II FPGA with 200~Msps 12 bit ADC.

\section{Characterization of calorimeter parameters}
\begin{figure}[h!]
\includegraphics[scale=0.20]{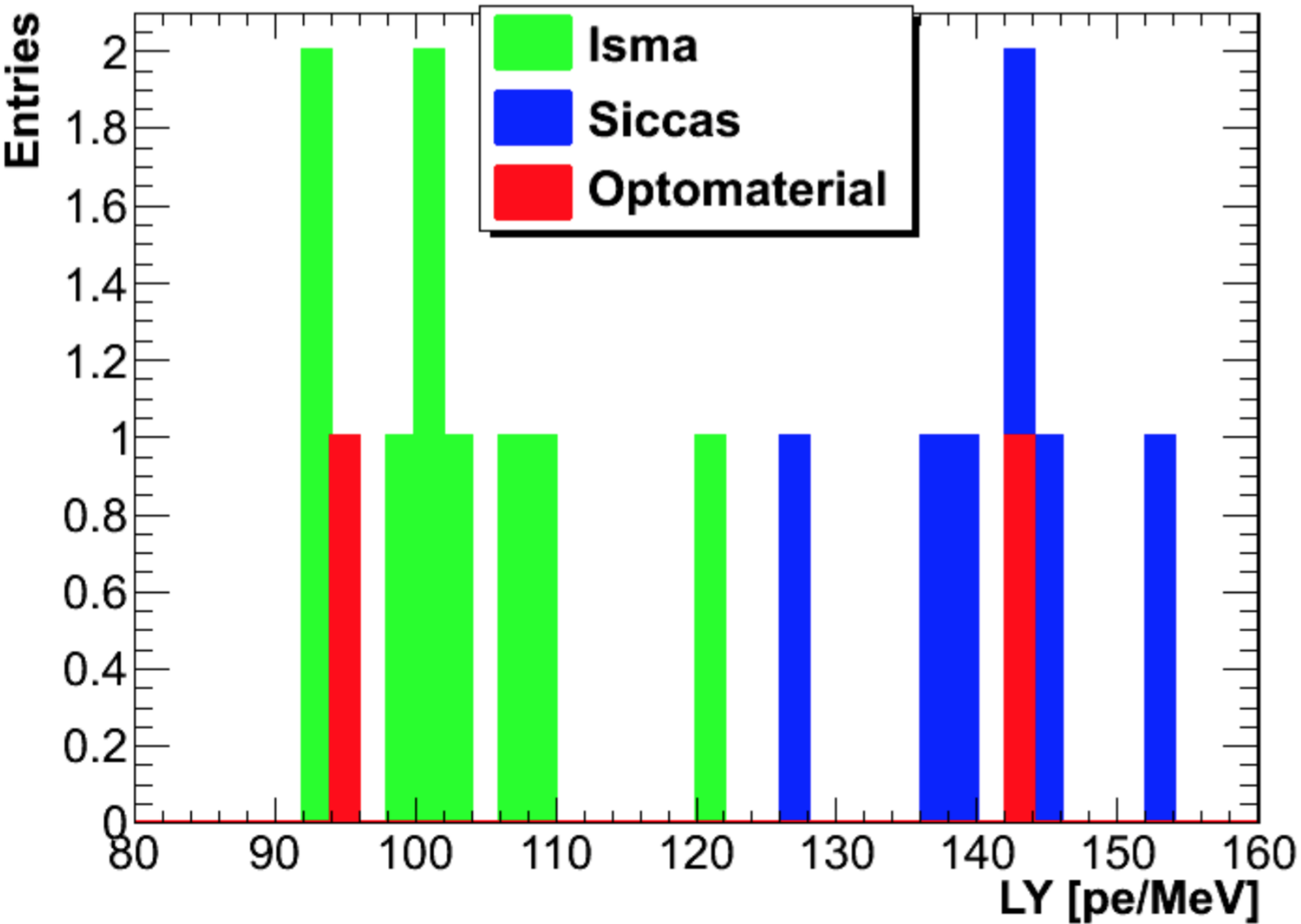}
\includegraphics[scale=0.21]{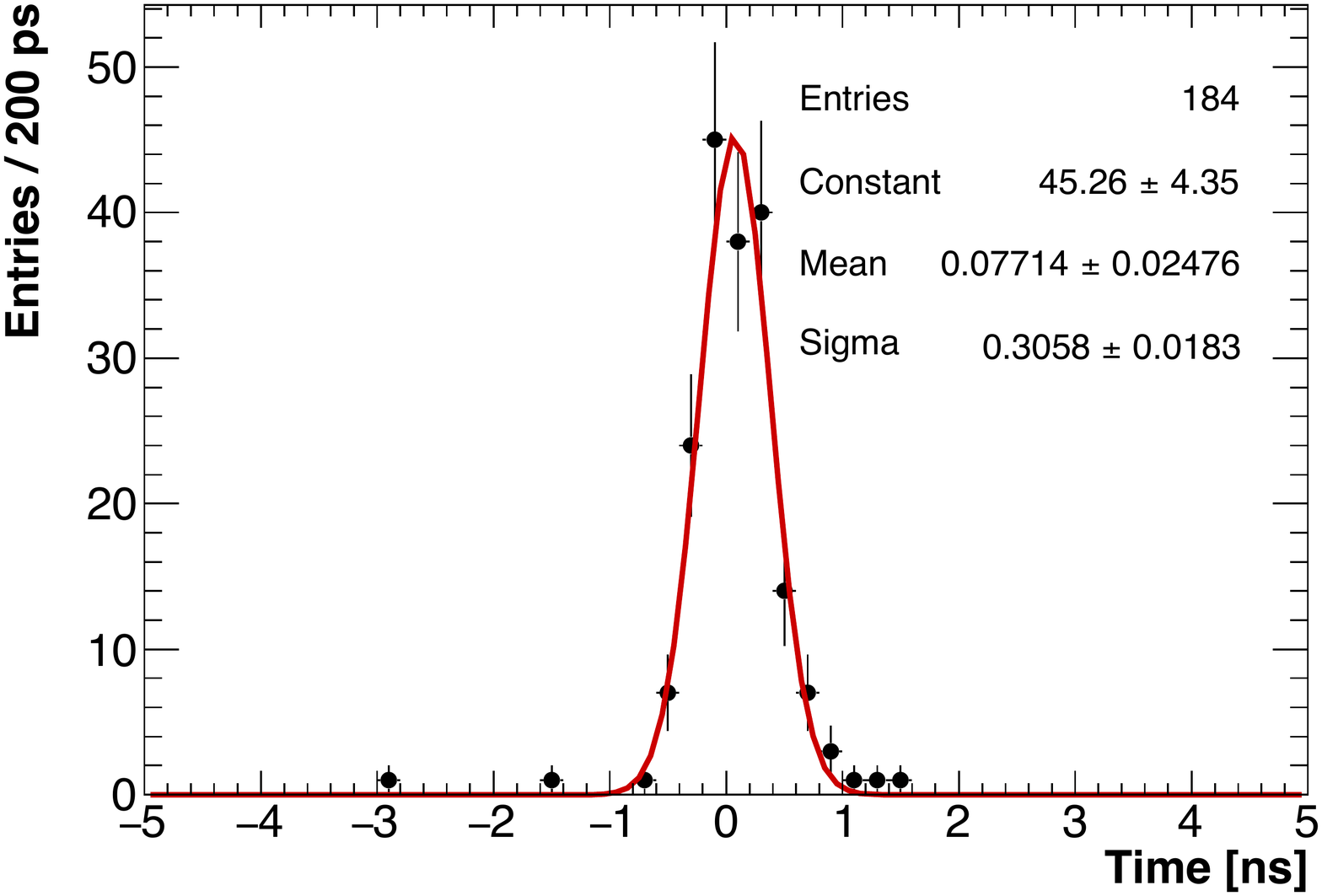}
\caption{\label{plot} Average LY of all the crystals tested (left) and time resolution of an Opto Materials crystal coupled with a single ($2 \times 3$) array of $6 \times 6$ mm$^2$ Hamamatsu SiPM (right).}
\end{figure}

Tests on CsI crystals have been performed with $^{22}$Na source for three different vendors: ISMA (Ukraine), SICCAS (China) and Opto Materials (Italy). 
All tested crystals show a good LY $\sim$~120 photoelectrons per MeV and a $\sim$~0.6\%/cm LRU when coupled with an UV-extended photomultiplier (PMT) and Tyvek wrapping (Fig.~\ref{plot}, left).
Exploiting cosmic rays and using a single ($2 \times 3$) array of $6 \times 6$ mm$^2$ Hamamatsu\cite{hama} SiPM as readout , we have evaluated also the time resolution, which is  $\sim$170 ps (@~$\sim$~22~MeV, energy deposited by a minimum ionizing particle in a CsI crystal) after subtracting the 255~ps of the trigger time resolution (Fig.~\ref{plot}, right).

Following calorimeter requirements, one important aspect to be considered is the radiation hardness. In this context, we have performed different tests both on crystals and SiPMs from different vendors.

Some crystals have been irradiated up to 900 Gy and to a neutron fluency up to $9 \times 10^{11}$~n$_{1MeV}$/cm$^2$. The ionization dose does not modify LRU while a 20\% reduction in LY has been observed at 900 Gy. Similarly, the neutron flux causes a 15\% LY deterioration. 
Moreover, it is important to control the noise induced by the instantaneous dose (2~rad/h) and thermal neutron flux (10~kHz/cm$^2$).
For this purpose, a crystal readout by a PMT has been irradiated in these conditions and the photocurrent has been recorded.
The energy equivalent noise, RIN, was derived as the standard deviation of the number of photoelectron, N, in a readout gate of 200~ns:
\begin{equation}
RIN = \frac{\sqrt{N}}{LY} (MeV)
\label{eq}
\end{equation}
We have measured the RIN from dose and thermal neutrons for crystals from the three vendors. Our results show the RIN from $\gamma$-ray in the hottest region to be around~300 keV. For thermal neutrons the RIN is much lower: 60-85~keV for a flux of 10$^4$~n/cm$^2$/s.

UV-extended SiPM, both Hamamatsu and FBK \cite{fbk} companies, have been irradiated with a dose up to 20~krad, which did not effect the leakage current. On the contrary, a current increase is clearly visible in all SiPMs when exposing the sensors to a total flux of $2.2 \times 10^{11}$~n/cm$^2$ (corresponding to 2.2 times the experiment lifetime)\cite{raffa_calor}: the leakage current of the Hamamatsu SiPM increased from 
$\sim16~\mu$A to $\sim2$~mA while the FBK one from  $\sim~21~\mu$A to $\sim~5$~mA. Even if the hall temperature was quite 
stable during irradiation the drop on the gain was mostly dominated by the temperature increase of the SiPMs.
To reduce it to acceptable value, we need to cool down all SiPM to a temperature of 0$^\circ$C. In order to do so, we will use a dedicated cooling station for the calorimeter, which is now under design.
\begin{figure}[h]
\includegraphics[scale=0.2]{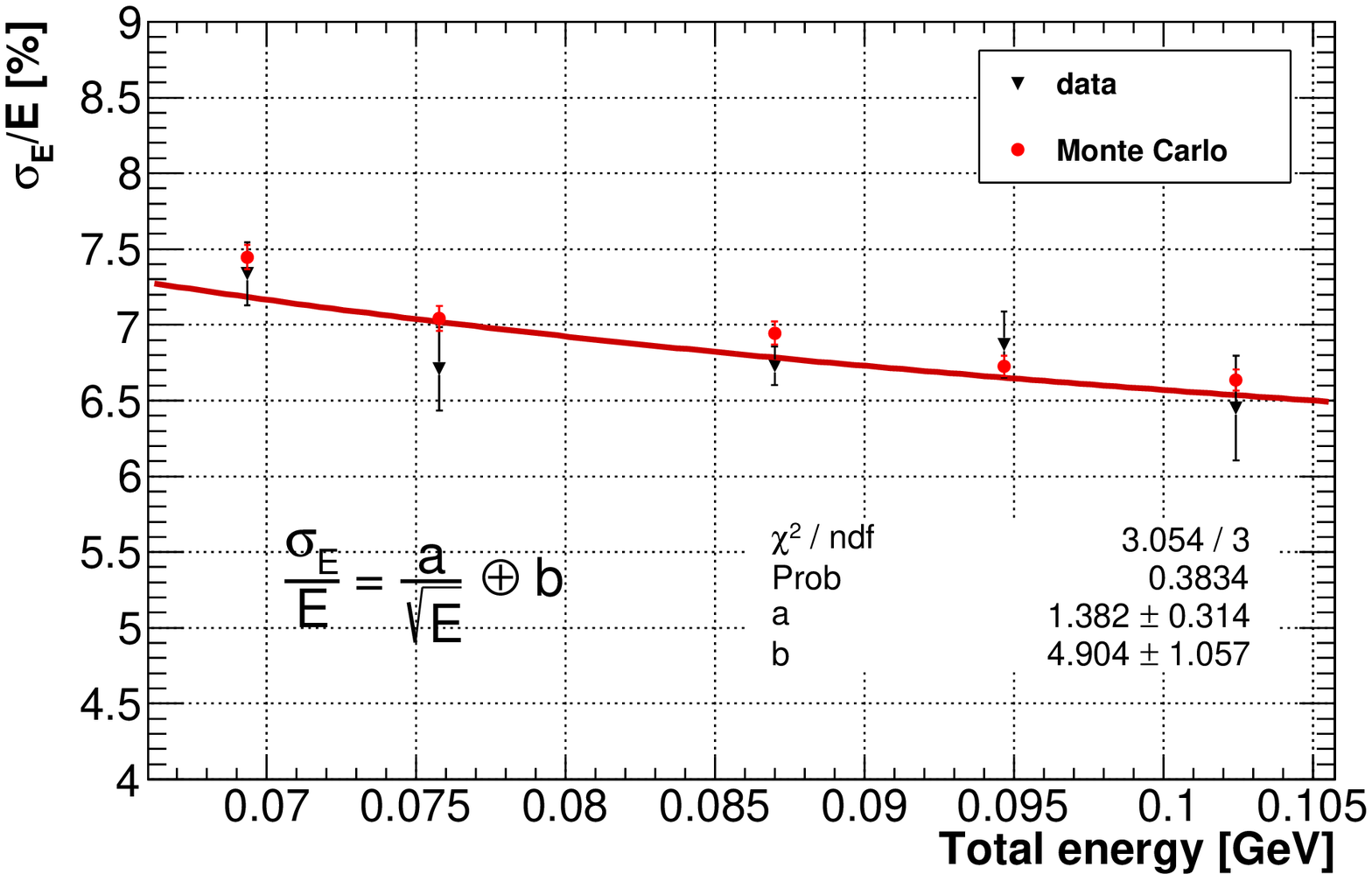}
\includegraphics[scale=0.193]{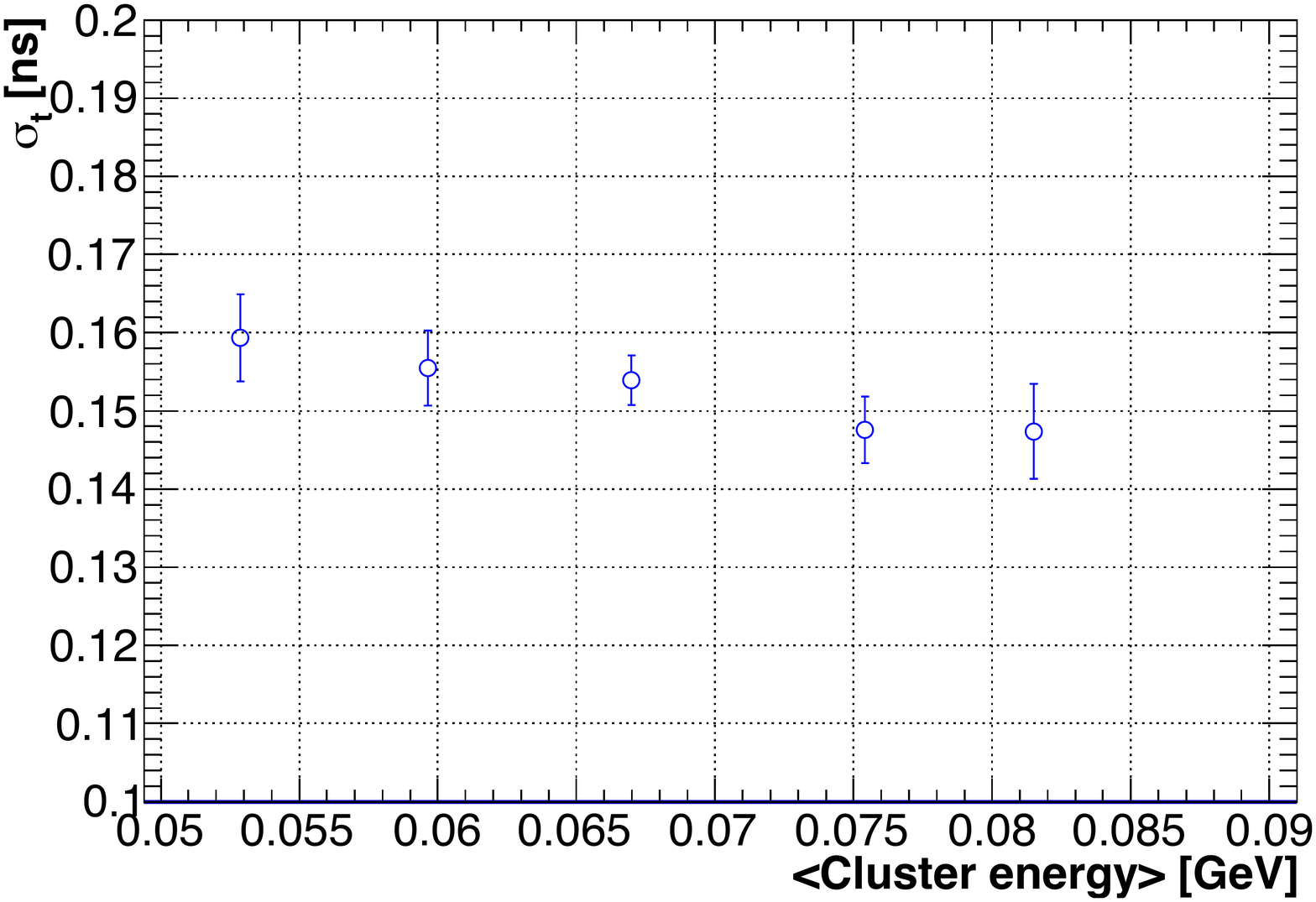}
\caption{\label{testbeam} Energy (left) and time (right) resolution at different electron beam energies of the $3\times3$ CsI matrix.}
\end{figure}

Finally, a small undoped CsI $3\times3$ matrix has been built and tested at the Frascati Beam Test Facility using electrons with energy between 80 and 120 MeV. Each crystal is read out using an array of sixteen ($3\times3$)~mm$^2$ Hamamatsu TSV SiPMs. During this test we measured a LY of 30 (20) pe/MeV with (without) optical grease with Tyvek wrapping. The measured time and energy resolution are 110~ps and 7\% respectively (Fig.~\ref{testbeam}).

These performance results, both of single crystals and of the small calorimeter prototype, are fully compatible with the requirements of the calorimeter. We are now preparing the international bid for the procurement of the pre-production and production crystals and sensors.

\section*{Acknowledgments}
Talk given at the 5th Young Researchers Workshop   
 "Physics Challenges in the LHC Era", 
 Frascati, May 9 and 12, 2016

 This work was supported by the EU Horizon 2020 Research and Innovation Programme under the Marie Sklodowska-Curie Grant Agreement No. 690835.


%
\end{document}